\documentclass[prd,twocolumn,showpacs,preprintnumbers,amsmath,amssymb,superscriptaddress,floatfix,nofootinbib]{revtex4}

\usepackage{graphicx}
\usepackage{amsmath}
\usepackage{amsfonts}
\usepackage{amssymb}
\usepackage{color}
\usepackage{multirow}
\usepackage[colorlinks, citecolor=blue,anchorcolor=red,menucolor=red,linkcolor=red,filecolor=red,runcolor=red,urlcolor=blue,frenchlinks=red]{hyperref}

\begin{document}

\title{The $\phi(2170)$ production in the process $\gamma p\to \eta \phi p$}

\author{Chen-Guang Zhao}
\affiliation{School of Physics and Engineering, Zhengzhou University, Zhengzhou, Henan 450001, China}

\author{Guan-Ying Wang}
\affiliation{School of Physics and Engineering, Zhengzhou University, Zhengzhou, Henan 450001, China}

\author{Guan-Nan Li}
\affiliation{School of Physics and Engineering, Zhengzhou University, Zhengzhou, Henan 450001, China}

\author{En Wang}\email{wangen@zzu.edu.cn}
\affiliation{School of Physics and Engineering, Zhengzhou University, Zhengzhou, Henan 450001, China}

\author{De-Min Li}\email{lidm@zzu.edu.cn}
\affiliation{School of Physics and Engineering, Zhengzhou University, Zhengzhou, Henan 450001, China}

\date{\today}

\begin{abstract}
We have studied the $\gamma p\to \eta \phi p$ reaction within the effective Lagrangian approach, by considering the contribution of the intermediate state $\phi(2170)$ production, and the background contributions of $t$-channel $\pi^0$ and $\eta$ mesons exchanges with the intermediate states $N$ and $N(1535)$. Our calculations show that there may be a peak, at least a bump structure around 2180 MeV associated to the resonance $\phi(2170)$ in the $\eta\phi$ mass distribution. We suggest to search for the resonance $\phi(2170)$ in this reaction, which would be helpful to shed light on its nature.

\end{abstract}
\maketitle

\section{INTRODUCTION}{\label{INTRODUCTION}}

The exotic hadrons beyond the conventional quark model are permitted to exist in the framework of the Quantum Chromodynamics. Some possible candidates for the exotic states have been accumulated experimentally~\cite{Guo:2017jvc,Klempt:2007cp,Ali:2017jda,Esposito:2016noz,Chen:2016qju,Lebed:2016hpi},  and there are many different explanations for their natures. Further study on the possible candidates for the exotic states is obviously needed both theoretically and experimentally.

The state $\phi(2170)$ [also denoted as $X(2170)$ or $Y(2170)$ in some literature] was observed by the Babar Collaboration via the initial-state-radiation process $e^+ e^- \to \gamma \phi f_0(980)$~\cite{Aubert:2006bu}, and later confirmed by Belle~\cite{Shen:2009zze}, BESII~\cite{Ablikim:2007ab}, and BESIII Collaborations~\cite{Ablikim:2014pfc, Ablikim:2017auj}. On the $\phi(2170)$ nature, various interpretations exist in the literature, such as the conventional strangeonium~\cite{Ding:2007pc,Wang:2012wa,Pang:2019ttv}, tetraquark state~\cite{Ali:2011qi,Chen:2018kuu,Wang:2006ri,Chen:2008ej,Ke:2018evd}, hybrid~\cite{Ding:2006ya,Page:1998gz}, $\Lambda \bar{\Lambda}$~\cite{Dong:2017rmg,Zhao:2013ffn}
 or $\phi f_0(980)$ molecule derived from full Faddeev equations with the $\phi K \bar{K}$ and coupled channels system (see also attempts to re-derive this picture with different approximations to the Faddeev equations in Refs.~~\cite{AlvarezRuso:2009xn,MartinezTorres:2008gy,MartinezTorres:2010ax}).
 The available experimental information of the $\phi(2710)$ is not enough to confirm or refute
one of the above interpretations. By now, all the experimental information about the $\phi(2170)$ was obtained from the $e^+e^-$ collision experiments. The information about the $\phi(2170)$ production in other processes will be helpful to shed light on its nature.

As we know, the associate production of hadrons by photon has been extensively studied since it provides an excellent
tool to learn details of the hadron spectrum~\cite{Xie:2013mua,Wang:2014jxb,Wang:2016dtb,Wang:2017hug}. The intense photon beams can be used to study the strangeonium-like states because of the strong affinity of the photon for $s\bar{s}$. For instance, the
strangeness $\phi(1019)$ and $\phi(1680)$ were observed in the photo-production reactions respectively in Refs.~\cite{Mibe:2005er,Seraydaryan:2013ija,Dey:2014tfa,Mizutani:2017wpg} and Refs.~\cite{Aston:1981tb,Atkinson:1984cs,Busenitz:1989gq}. The observation of the $\phi(2170)$  in $\phi f_0(980)$ indicates that the $\phi(2170)$ has a substantial $s\bar{s}$ component. Thus, the photo-production reaction could be also suitable to study the resonance $\phi(2170)$.

It should be pointed out that, in Fig.~25 of Ref.~\cite{Busenitz:1989gq}, which corresponds to the $d\sigma/dM_{K\bar K}$
 distribution for the reaction $\gamma N \to K^+ K^-N$ measured by the Omega Photon Collaboration, besides the peak of the $\phi(1680)$, another enhancement around 2150~MeV exists, which could be associated to a resonance with the same quantum numbers as $\phi(1680)$, i.e. $J^{PC}=1^{--}$. In addition, BESIII also observed a resonant structure around 2200~MeV at the cross section of $e^+e^- \to K^+ K^-$ reaction~\cite{Ablikim:2018iyx}.  The $K\bar K$ channel is expected to be one of the main decay channels of the  $\phi(2170)$~\cite{Ding:2007pc,Wang:2012wa,Pang:2019ttv}. Thus, it is natural to associate this enhancement structure around 2150~MeV in Fig.~25 of Ref.~\cite{Busenitz:1989gq} to the $\phi(2170)$, which implies that the $\phi(2170)$ photo-production should be accessible experimentally.

Since the $\phi(2170)$ was observed in the $\phi f_0(980)$ channel~\cite{Aubert:2006bu}, it would be straightforward to search for the resonance $\phi(2170)$ in the reaction of $\gamma p \to \phi f_0(980) p$. However, one would face the possible mixing between the $\phi(2170)$ and the threshold effect in $\gamma p \to \phi f_0(980) p$ reaction because the mass of $\phi(2170)$ is close to the threshold of $\phi f_0(980)$.
The measurement of $\Gamma(\phi(2170)\rightarrow\eta\phi)/\Gamma(\phi(2170)\rightarrow\phi f_0(980))=(1.7\pm 0.7\pm 1.3)/(2.5\pm 0.8\pm 0.4)$~\cite{PDG2018}
indicates that the coupling of the $\phi(2170)$ to the $\phi f_0(980)$ channel is of the same order of magnitude as its coupling to $\eta \phi$ channel, which suggests that the $\phi(2170)$ has a sizeable coupling to the $\eta\phi$ channel. Also, the threshold of $\eta \phi$ is about 1570~MeV, thus the signal of the $\phi(2170)$ can not be misidentified with the threshold effect. These factors encourage us to study the  $\phi(2170)$ production in the reaction of $\gamma p \to \eta \phi p$ within the effective Lagrangian approach. Our main purpose is to propose a possible process of searching for the $\phi(2170)$ resonance.

This paper is organized as follows. In Sec.~\ref{sec:formalism}, we present the mechanisms of the reaction  $\gamma p \to \eta \phi p$, our results and discussions are given in Sec.~\ref{sec:results}. Finally, a short summary is given in Sec.~\ref{sec:conclusion}.

\section{FORMALISMS}{\label{sec:formalism}}
\begin{figure*}[t]
  \centering
  \includegraphics[width=0.8\textwidth]{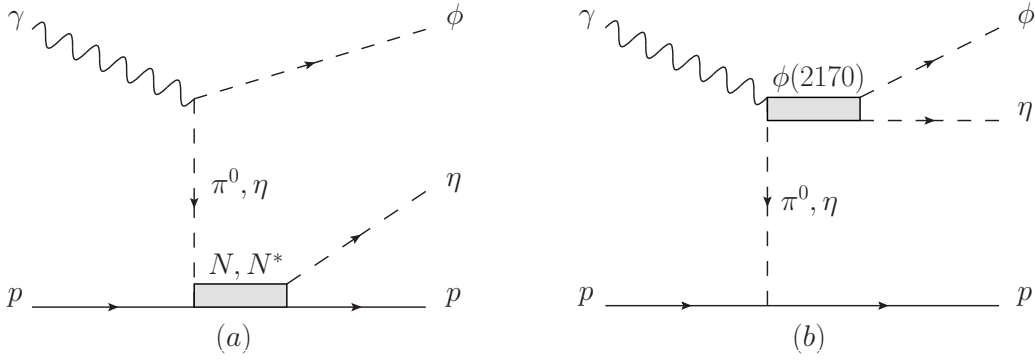}
  \caption{Feynman diagrams for the $\gamma p\to\eta \phi p$ reaction. (a) the contribution of the $t$-channel $\pi^0$ and $\eta$ exchanges with the intermediate states $N$ and $N^*$. (b) the contribution of the  intermediate states $\phi(2170)$ production.}
  \label{fig:feyn}
\end{figure*}

In this section, we will present the mechanisms for the reaction,
\begin{equation}
\gamma (p_1, s_1)+ p(p_2,s_2) \to \phi (p_3, s_3) +\eta (p_4) + p (p_5,s_5),
\end{equation}
by considering the tree level diagram as depicted in Fig.~\ref{fig:feyn}. $p_i\;(i=1,2,3,4,5)$ are the four-momenta for photon, initial proton, $\phi$, $\eta$, and outgoing proton, respectively, and $s_i(i=1,2,3,5)$ correspond to the polarizations of photon, initial proton, $\phi$, and outgoing proton, respectively. We consider the background contribution of the $t$-channel $\pi^0$ and $\eta$ exchanges with the final state $\eta p$ producing through the intermediate states $N$ and $N^*$, as shown in Fig.~\ref{fig:feyn}(a). The $\phi(2170)$ can be directly produced by $t$-channel $\pi^0$ and $\eta$ exchanges, and then decays to $\eta\phi$, which is shown in Fig.~\ref{fig:feyn}(b).

We use the commonly employed Lagrangian densities for $\pi N N$ and $\eta N N$ as follows~\cite{Liang:2004sd,Lu:2014yba}:
\begin{align}
\mathcal{L}_{\pi N N} \;&=\; -\frac{g_{\pi N N}}{2m_N} \bar{N} \gamma_5 \gamma_{\mu} \vec{\tau} \cdot \partial^{\mu} \vec{\pi} N  ,\\
\mathcal{L}_{\eta N N} \;&=\; -\frac{g_{\eta N N}}{2m_N} \bar{N} \gamma_5 \gamma_{\mu} \partial^{\mu} \eta N  ,
\end{align}
with the coupling constants $g_{\pi N N}=13.45$ and $g_{\eta N N}=2.24$, taken from Refs.~\cite{Xie:2007qt,Liang:2004sd,Lu:2014yba}.
For the $\gamma\phi \pi$ and $\gamma\phi \eta$ couplings, we take the interaction Lagrangian densities as used in Refs.~\cite{Ryu:2012tw},
\begin{align}
\mathcal{L}_{\gamma \phi \pi} \;&=\; \frac{e}{m_\phi} g_{\gamma \phi \pi} \epsilon^{\mu \nu \alpha \beta} \partial_{\mu} \phi_{\nu} \partial_{\alpha} A_{\beta} \pi
,\\
\mathcal{L}_{\gamma \phi \eta} \;&=\; \frac{e}{m_\phi} g_{\gamma \phi \eta} \epsilon^{\mu \nu \alpha \beta} \partial_{\mu} \phi_{\nu} \partial_{\alpha} A_{\beta}
\eta ,
\end{align}
where $e$ is the electron charge, $\phi_\nu$, $A_\beta$, $\pi$, and $\eta$ are the fields of $\phi$, photon, $\pi$, and $\eta$, respectively.

For the intermediate nucleon excited resonances, we only take into account the contribution of the resonances $N(1535)$, since the $\eta p$ production from the intermediate resonance $N(1535)$  plays the dominant role near the threshold, as discussed in Refs.~\cite{Lu:2014yba,Fan:2019lwc} where more nucleon excited resonances were considered.
We also need the interaction Lagrangian densities involving $N(1535)$ (marked as $N^*$)~\cite{Zou:2002yy},
\begin{align}
\mathcal{L}_{\pi N N^*} \;&=\; i g_{\pi N N^*} \bar{N^*} \vec{\tau} \cdot \vec{\pi} N + {\rm H.c.}  ,\\
\mathcal{L}_{\eta N N^*} \;&=\; i g_{\eta N N^*} \bar{N^*} \eta N + {\rm H.c.}  .
\end{align}

For the intermediate state $\phi(2170)$ production of Fig.~\ref{fig:feyn}(b), we only consider the $t$-channel $\eta$ exchange, because there is no any information about the radiative decays of the $\phi(2170)$, and the process $\phi(2170)\to \pi\gamma$ should be suppressed due to the isospin breaking effect. More comprehensive mechanism can be considered if more experimental information about this reaction is available.
The interaction Lagrangian densities involving $\phi(2170)(\equiv  \phi^*)$ meson~\cite{Piotrowska:2017rgt} are,
\begin{align}
\mathcal{L}_{\phi^* \gamma \eta} \;&=\; \frac{e}{m_{\phi^*}} g_{\phi^* \gamma \eta} \epsilon^{\mu \nu \alpha \beta} \partial_{\alpha} \phi^*_{\beta} \partial_{\mu}
A_{\nu} \eta ,\\
\mathcal{L}_{\phi^* \phi \eta} \;&=\; \frac{g_{\phi^* \phi \eta}}{m_{\phi^*}}  \epsilon^{\mu \nu \alpha \beta} \partial_{\alpha} \phi^*_{\beta} \partial_{\mu}
\phi_{\nu} \eta.
\end{align}

The coupling constants in the above Lagrangian densities can be determined from the partial decay widths,
\begin{align}
\Gamma[\phi\to\pi\gamma] \;&=\; \frac{e^2 g^2_{\phi \gamma \pi}}{12\pi} \frac{|\vec{p}_{\pi\gamma}|^3}{m^2_\phi},\\
\Gamma[\phi\to\eta\gamma] \;&=\; \frac{e^2 g^2_{\phi \gamma \eta}}{12\pi} \frac{|\vec{p}_{\eta\gamma}|^3}{m^2_\phi},
\end{align}
for the $\phi$ meson,
\begin{align}
\Gamma[N^*\to N\pi] \;&=\; \frac{3g^2_{\pi N N^*}}{4\pi} \frac{E_N+m_N}{m_{N^*}} |\vec{p}_{N\pi}|,\\
\Gamma[N^*\to N\eta] \;&=\; \frac{g^2_{\eta N N^*}}{4\pi} \frac{E_N+m_N}{m_{N^*}} |\vec{p}_{N\eta}|,
\end{align}
for $J^P (\frac{1}{2}^-)$ nucleon excited resonance $N(1535)$,
\begin{align}
\Gamma[\phi^*\to\phi\eta] \;&=\; \frac{g^2_{\phi^* \phi \eta}}{12\pi} \frac{|\vec{p}_{\eta\phi}|^3}{m^2_{\phi^*}},\\
\Gamma[\phi^*\to\eta\gamma] \;&=\; \frac{e^2 g^2_{\phi^* \gamma \eta}}{12\pi} \frac{|\vec{p}_{\eta\gamma}|^3}{m^2_{\phi^*}},
\end{align}
for the $\phi(2170)$, with
\begin{align}
|\vec{p}_{f_1 f_2}| \;&=\; \frac{\lambda^{\frac{1}{2}}(m^2_i,m^2_{f_1},m^2_{f_2})}{2m_i},
\end{align}
where $m_i$ denotes the mass of the initial state, $m_{f_1}$ and $m_{f_2}$ are the masses of two final states, and $\lambda$ is the K\"{a}llen function with
$\lambda(x,y,z)=(x-y-z)^2-4yz$.

At present, there is no information about $\Gamma(\phi(2170)\rightarrow \eta\phi)$ and $\Gamma(\phi(2170)\rightarrow\gamma\eta)$ experimentally. Model predictions on $\Gamma(\phi(2170)\to \eta \phi)$ lie a region from about 1 to 20 MeV~\cite{Ding:2006ya,Wang:2012wa,Pang:2019ttv,Barnes:2002mu}. As far as we know, the radiative decays of $\phi(2170)$ have so far never been investigated in any models. We can roughly estimate the value of $\Gamma(\phi(2170)\rightarrow \gamma\eta)$ in the quark model based on the $M1$ radiative partial width between the $v=n^{2S+1}L_J$ and $v^\prime = n^{\prime 2S+1}L^\prime_{J^\prime}$ states~\cite{Li:2010vx,Lu:2016bbk}
\begin{eqnarray}
&&\Gamma_{M1}(v \to v^\prime + \gamma) \nonumber\\
&&= \frac{\alpha e^{\prime 2}_Q}{3}\frac{2J^\prime +1}{2L+1}\delta_{LL^\prime}\delta_{SS^\prime \pm 1}|<v^\prime|j_0(\frac{E_\gamma r}{2})|v>|^2 \frac{E^3_\gamma E_f}{M_i},\nonumber\\
\end{eqnarray}
where $e^\prime _Q=\frac{m_1 Q_2 +m_2 Q_1}{m_1m_2}$, $m_1$ and $m_2$ are the quark 1 and 2 masses, respectively,  $Q_1$ and $Q_2$ stand for the quark $1$ and $2$ charges in units of $|e|$, respectively. $\alpha = 1/137$ is the fine-structure constant, $E_\gamma$ is the photon energy, $E_f$ is the energy of final  meson, $M_i$ is the mass of initial state.
In the picture of the $\phi$, $\phi(1680)$, and $\phi(2170)$ as the $1^3S_1$, $2^3S_1$, and $3^3S_1$ $s\bar{s}$, respectively, employing the wavefunctions obtained from the Godfrey-Isgur relativized quark model~\cite{Godfrey:1985xj}, we have
\begin{eqnarray}
\frac{\Gamma(\phi(1680)\rightarrow\gamma\eta)}{\Gamma(\phi\rightarrow\gamma\eta)}\simeq 1.1,\\
\frac{\Gamma(\phi(2170)\rightarrow\gamma\eta)}{\Gamma(\phi\rightarrow\gamma\eta)}\simeq 3,
\end{eqnarray}
which lead to $\Gamma(\phi(1680)\rightarrow\gamma\eta)\simeq 0.06$ MeV and $\Gamma(\phi(2170)\rightarrow\gamma\eta)\simeq 0.17$ MeV with the central value of  Br$(\phi\rightarrow\gamma\eta)=1.303\%$~\cite{PDG2018}. The $\Gamma(\phi(1680)\rightarrow\gamma\eta)$ has been predicted to be about 0.09~MeV in a quark model with Gaussian wavefunctions~\cite{Close:2002ky} or $0.14\pm 0.09$~MeV in an effective relativistic quantum field theoretical model based on flavor symmetry~\cite{Piotrowska:2017rgt}. Our predicted $\Gamma(\phi(1680)\rightarrow\gamma\eta)\simeq 0.06$ MeV is smaller than the result of Ref.~\cite{Close:2002ky} and close to the lower limit of the result of Ref.~\cite{Piotrowska:2017rgt}, thus, one can conjecture that $\Gamma(\phi(2170)\rightarrow\gamma\eta)\simeq 0.17$ MeV maybe also correspond to the lower limit of other model predictions. In the present work, we shall take $\Gamma(\phi(2170)\rightarrow\eta\phi)=1$ MeV and $\Gamma(\phi(2170)\rightarrow\gamma\eta)\simeq 0.17$ MeV. We want to test whether the signal of $\phi(2170)$ is observable in $\gamma p\rightarrow\eta\phi p$  with the lower limit of $g_{\phi^*\phi\eta}$ and $g_{\phi^*\gamma \eta}$.  The obtained results for the coupling constants are listed in Table~\ref{tab:coupling}.

\begin{table}
\begin{center}
\caption{ \label{tab:coupling} Model parameters used in the present work, the masses, widths, and branching ratios are taken from Particle Data Group~\cite{PDG2018}.}
\footnotesize
\begin{tabular}{cccccc}
\hline\hline
  State         &Mass &Width      &Decay          &Adopted                &$g^2/4\pi$             \\
                &(MeV)&(MeV)      &channel        &branching ratio     &                            \\ \hline
  $\phi$        &1019&4.25       &$\gamma\pi$    &$1.3\times 10^{-3}$   &$1.60\times 10^{-3}$   \\
                &&           &$\gamma\eta$   &$1.3\times 10^{-2}$   &$3.97\times 10^{-2}$   \\
  $N(1535)$    &1535&150        &$N\pi$         &0.42                   &$3.43\times 10^{-2}$   \\
                &&           &$N\eta$        &0.42                   &0.28                   \\
  $\phi(2170)$  &2188&83         &$\gamma\eta$   &   -                    & $2.41\times 10^{-2}$                      \\
                &&           &$\phi\eta$                                                    & - & $3.59 \times 10^{-2}$\\
\hline\hline
\end{tabular}
\end{center}
\end{table}

Since the hadrons are not point-like particles, the form factors are also needed. We adopt the dipole form factor,
\begin{align}
\mathcal{F}_M = \left(\frac{\Lambda^2_M-M^2}{\Lambda^2_M- q^2}\right)^2, \label{eq:FFM}
\end{align}
 for exchanged mesons~\cite{Xie:2007qt,Oh:2007jd,Gao:2010hy,Wang:2017hug}, and
 \begin{align}
\mathcal{F}_B = \frac{\Lambda^4_B}{\Lambda^4_B+(q^2-M^2)^2},
\end{align}
for the exchanged baryons~\cite{Feuster:1997pq,Xie:2013mua,Wang:2014jxb,Wang:2017hug},
where the $q$ and $M$ are the four-momentum and the mass of the exchanged hadron, respectively.
Indeed, the values of the cutoff parameters can be directly related to the hadron size. Since the question of hadron size is still open, the cutoff parameters are usually adjusted to the related experimental measurement~\cite{Xie:2015zga,Wang:2016fhj}. The typical value of the cut-off $\Lambda$ in the born potential is in the region of $1.0\sim 2.0$~GeV.
In our present calculation, we use the cut-off parameters $\Lambda_\pi=\Lambda_\eta=1.5$~GeV, and $\Lambda_N=\Lambda_{N^*}=2.0$ GeV for $N$ and $N(1535)$ as used in Refs.~\cite{Xie:2007qt,Lu:2014yba}.
For the one of $\phi(2170)$, we take $\Lambda_{\phi^*}=1.0$~GeV.

The propagators used in our calculation are
\begin{equation}
G_{\pi,\eta}(q) = \frac{i}{q^2 - M^2 },
\end{equation}
for exchanged $\pi$, $\eta$ mesons,
\begin{equation}
G_{N,N^*}(q) = i \frac{\not\!{q} + M }{q^2 - M^2 + i M \Gamma},
\end{equation}
for the propagator of spin-1/2 baryon,
and
\begin{equation}
G_{\phi^*}^{\mu \nu}(q) = i \frac{g^{\mu \nu} - q^{\mu} q^{\nu} / {M^2} }{q^2 - M^2 + i M \Gamma}, \label{eq:BW2170}
\end{equation}
for the propagator of $\phi(2170)$,
where $q$, $M$, and $\Gamma$ stand for the four-momentum, mass, and total width of the resonance, respectively. Indeed, as we discussed in the introduction, there are many interpretations on the nature of $\phi(2170)$, and all of them give rise to the similar structures as the Breit-Wigner form of Eq.~(\ref{eq:BW2170}).

With the above effective Lagrangian densities, the scattering amplitudes for the reaction $\gamma p\to\eta\phi p$ can be obtained straightforwardly as follows,
\begin{eqnarray}
\mathcal{M}^\pi_{N} &=& \frac{i e g_{\pi N N} g_{\eta N N} g_{\gamma \phi \pi}}{(2M_N)^2 m_\phi} \mathcal{F}(q^2_\pi,M_\pi^2) \mathcal{F}(q^2_N,M_N^2)  \nonumber
\\&& \times\bar{u}(p_5,s_5) \gamma_5 \not\!{p_4} G_{N}(q_N) \gamma_5 \not\!{q_\pi} u(p_2,s_2) G_\pi(q_\pi)\nonumber
\\&& \times\epsilon^{\mu \nu \alpha \beta} p_{3\mu} \varepsilon^*_{\nu}(p_3,s_3) p_{1\alpha} \varepsilon_{\beta}(p_1,s_1),
\end{eqnarray}
for the contribution of the $t$-channel $\pi^0$ exchange with the intermediate nucleon, and
\begin{eqnarray}
\mathcal{M}^\pi_{N^*} &=& -\frac{i e g_{\pi N N^*} g_{\eta N N^*} g_{\gamma \phi \pi}}{m_\phi} \mathcal{F}(q^2_\pi,M_\pi^2) \mathcal{F}(q^2_{N^*},M_{N^*}^2)  \nonumber
\\&& \times\bar{u}(p_5,s_5) G_{N^*}(q_{N^*}) u(p_2,s_2) G_\pi(q_\pi)\nonumber
\\&& \times\epsilon^{\mu \nu \alpha \beta} p_{3\mu} \varepsilon^*_{\nu}(p_3,s_3) p_{1\alpha} \varepsilon_{\beta}(p_1,s_1),
\end{eqnarray}
for the contribution of the $t$-channel $\pi^0$ exchange with the intermediate $N(1535)$. The amplitudes $\mathcal{M}^\eta_{N}$ and $\mathcal{M}^\eta_{N^*}$ due to the $\eta$ exchange can be obtained from $\mathcal{M}^\eta_{N}$ and $\mathcal{M}^\eta_{N^*}$ by replacing $\pi$ by $\eta$.

The amplitude of the intermediate $\phi(2170)$ production can be written as,
\begin{eqnarray}
\mathcal{M}_{\phi^*} &=& -\frac{i e g_{\eta N N} g_{\phi^* \phi \eta} g_{\phi^* \gamma \eta}}{2M_N m_\phi^2} \mathcal{F}(q^2_\eta,M_\eta^2) \mathcal{F}(q^2_{\phi^*
},M_{\phi^*}^2)  \nonumber
\\&& \times\bar{u}(p_5,s_5) \gamma_5 \not\!{k_\eta}  u(p_2,s_2) G_\eta(q_\eta)\nonumber
\\&& \times\epsilon^{\mu \nu \alpha \beta} p_{3\mu} \varepsilon^*_{\nu}(p_3,s_3) p_{\phi^*\alpha} G_{\phi^* (\beta\rho)}(q_{\phi^*}) \nonumber
\\&& \times\epsilon^{\rho \sigma \delta \lambda} p_{\phi^*\sigma} p_{1\delta} \varepsilon_{\lambda}(p_1,s_1). \label{eq:ampphistar}
\end{eqnarray}
$q_\pi = p_1-p_3$ is the four-momentum for the exchanged $\pi$ meson, and $q_R = p_4 + p_5$ is the
four-momentum for intermediate $N$ and $N(1535)$. It is easy to show that the above amplitudes respect the gauge invariance~\cite{Lu:2014yba}.

Then the differential cross section for the reaction $\gamma p\to \eta\phi p$ can be expressed as,
\begin{align}
\qquad & d\sigma(\gamma p\to \eta\phi p) = \frac{1}{8E_\gamma} \bar\sum |\mathcal{M}_{\rm total}|^2 \times \nonumber \\
& \frac{d^3p_3}{2E_3} \frac{d^3p_4}{2E_4} \frac{m_p d^3p_5}{E_5} \delta^4(p_1+p_2-p_3-p_4-p_5),
\end{align}
with
\begin{eqnarray}
\mathcal{M}_{\rm total} = \mathcal{M}^\pi_N+\mathcal{M}^\pi_{N^*}+\mathcal{M}^\eta_N+\mathcal{M}^\eta_{N^*}+\mathcal{M}_{\phi^*},
\end{eqnarray}
where $E_3$, $E_4$, and $E_5$ are the energies of the $\phi$, $\eta$, and outgoing proton, respectively, and $E_\gamma$ is the photon energy in the laboratory frame.

\section{RESULT AND DISCUSSION}{\label{sec:results}}
With the above formalisms, we calculate the total and differential cross sections for the $\gamma p\to\eta \phi p$ reaction by using a Monte Carlo multi-particle phase space integration program used in Refs.~\cite{Lu:2014yba,Xie:2014kja}.

\begin{figure}[h]
\centering
\includegraphics[width=0.4\textwidth]{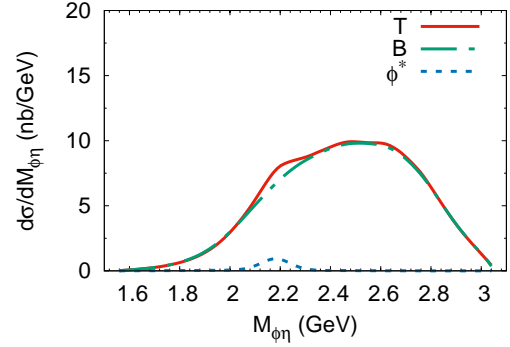}
 \caption{The $\eta\phi$ mass distribution of the $\gamma p\to\eta\phi p$ reaction with $E_\gamma=8$~GeV. The curves labeled as `B' and `$\phi^*$'stand for the contributions of the background and the intermediate $\phi(2170)$ production, respectively. The curve labeled as `T' corresponds to the total contributions.}
  \label{fig:dcs}
  \end{figure}

  \begin{figure}[h]
  \centering
  \includegraphics[width=0.4\textwidth]{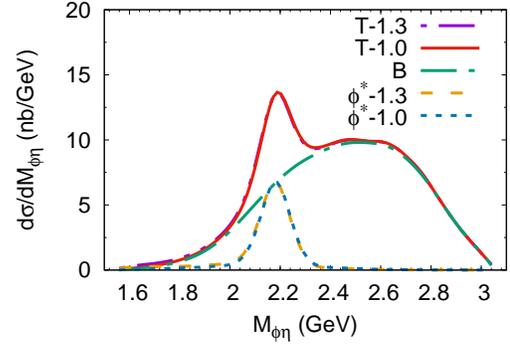}
\caption{The $\eta\phi$ mass distribution of the $\gamma p\to\eta\phi p$ reaction with $E_\gamma=8$~GeV in the presence of $\Gamma(\phi(2170)\rightarrow\eta\phi)\simeq6.6$~MeV. The curves labeled as `T-1.0' (`T-1.3') and `$\phi^*$-1.0' (`$\phi^*$-1.3') correspond to the contributions of the full model and the intermediate $\phi(2170)$ with the $\Lambda_{\phi^*}=1.0 (1.3)$~GeV, respectively.}
  \label{fig:dcsx}
 \end{figure}

The $\eta\phi$ mass distribution of the $\gamma p\to \eta\phi p$ reaction with $E_\gamma=8$~MeV is shown in Fig.~\ref{fig:dcs}.
  It should be pointed out that for the photo-production, there is a contribution from Pomeron exchange, whose effect is dominant at large center-of-mass energy and forward angle. However, in this paper only the $\eta\phi$ mass distribution is relevant to the signal of $\phi(2170)$, and the comprehensive mechanism involved the Pomeron exchange dose not change too much the shape of the $\eta\phi$ mass distribution.
 As we can see, there is a bump structure around 2180~MeV, which is associated to the resonance $\phi(2170)$.  Our model prediction is based on the lower limit of $g_{\phi^*\phi\eta}$ and $g_{\phi^*\gamma \eta}$. One can expect that if a larger value for the product of $g_{\phi^*\phi\eta}g_{\phi^*\gamma \eta}$ is used, the signal of the $\phi(2170)$ would be more clear. For example, if we take $\Gamma(\phi(2170)\rightarrow\eta\phi)\simeq 6.6$ MeV expected by the recent quark model calculations~\cite{Pang:2019ttv}, one can find a significant peak around 2180 MeV as shown in Fig.~\ref{fig:dcsx} (see the curve labeled as `T-1.0').

In addition to the couplings related to the $\phi(2170)$ state, the $\phi(2170)$ form factor of Eq.~(\ref{eq:FFM}) depended on the $\Lambda_{\phi^*}$, would affect the numerical results. We also show the results with cutoff $\Lambda_{\phi^*}=1.3$~GeV in Fig.~\ref{fig:dcsx}, where the curves labeled as `T=1.3' and `$\phi^*$-1.3' correspond to the contributions of the full model and the intermediate $\phi(2170)$, respectively. It is found  that the results for $\Lambda_{\phi^*}=1.3$~GeV have little change compared with those for $\Lambda_{\phi^*}=1.0$~GeV. The exact value of $\Lambda_{\phi^*}$ can be extracted from the experimental measurement in future.

\begin{figure}[h]
  \centering
  \includegraphics[width=0.4\textwidth]{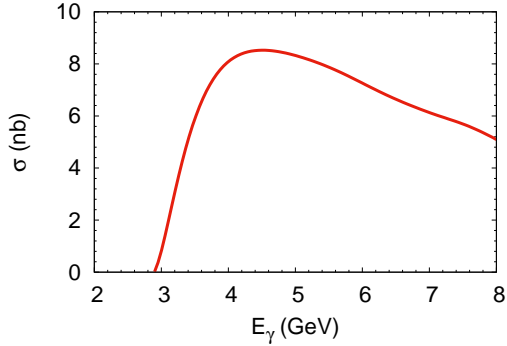}
  \caption{Total cross section of the $\gamma p\to\phi\eta p$ reaction.}
  \label{fig:tcs}
\end{figure}
In Fig.~\ref{fig:tcs}, we show the total cross section of the $\gamma p\to\eta\phi p$ reaction with the parameters listed in Table~\ref{tab:coupling} . Very recently, the reaction of $\gamma p\to\eta \phi p$ is also suggested to study the nucleon resonances production in Ref.~\cite{Fan:2019lwc} where the total cross section at $E_\gamma=3.8$~GeV is around $8\sim 10$~nb, which is consistent with our prediction.


%
%

 Finally, it should be noted that the GlueX Collaboration has proposed to search for the $\phi(2170)$ in the photoproduction~\cite{AlekSejevs:2013mkl}, and the $\gamma p \to \eta\phi p$ reaction has been selected as a particularly suitable process to search for strangeonium states by the CLAS12 Collaboration~\cite{Filippi:2015wea}. Our predictions should be useful for the future experimental study.

\section{Conclusions}
\label{sec:conclusion}
Motivated by the small enhancement around $2150$~MeV in the $K^+K^-$ mass distribution of the $\gamma p\to K^+K^-p$ reaction measured by Omega Photon Collaboration, and the clues that the branching ratio Br$(\phi(2170)\to \eta \phi)$ is of the same order as Br$(\phi(2170)\to \phi f_0(980))$, we propose to search for the resonance $\phi(2170)$ in the $\gamma p\to\eta\phi p$ reaction.

Our calculations show that there will be a peak, at least a bump structure around 2180~MeV in the $\eta\phi$ mass distribution of $\gamma p\to\phi\eta p$ reaction, the The magnitude of our total cross section is consistent with the prediction of Ref.~\cite{Fan:2019lwc} where the same reaction is used to study the nucleon resonances. We suggest our experimental colleagues to search for the resonance $\phi(2170)$ in the $\gamma p\to\eta\phi p$ reaction, which would be helpful to
shed light on its nature.

\section*{Acknowledgements}

We warmly thank Ju-Jun Xie, Qi-Fang ~L\"{u}, and Wen-Biao Yan for fruitful discussions. This work is partly supported by the National Natural
Science Foundation of China under Grant Nos. 11505158, 11605158. It is also supported by the Academic Improvement Project of Zhengzhou University.

\end{document}